\documentclass[journal,onecolumn,11pt]{IEEEtran}

\usepackage{graphicx,epsfig,subfigure}
\usepackage{times}
\usepackage{amsmath,amssymb,latexsym}
\usepackage{setspace}
\usepackage{longtable}
\usepackage{multicol}
\usepackage{cite}
\usepackage{color}
\usepackage{psfrag}

\newtheorem{theorem}{Theorem}
\newtheorem{corollary}{Corollary}

\providecommand{\E}{{\sf E}} 
\providecommand{\Cov}{{\sf cov}}

 \providecommand{\Av}{\mathbf{A}}

\providecommand{\Ac}{{\mathcal A}} \providecommand{\Cc}{{\mathcal C}}
 
 \providecommand{\Lc}{{\mathcal L}}

\providecommand{\Nc}{{\mathcal N}} 
\providecommand{\Pc}{{\mathcal P}} \providecommand{\Rc}{{\mathcal R}}

\providecommand{\hu}{\underline{h}} \providecommand{\gu}{\underline{g}}

\begin{document}
\doublespace

\title{Fading Cognitive Multiple-Access Channels With Confidential Messages}
\IEEEoverridecommandlockouts

\author{Ruoheng Liu, Yingbin Liang and H. Vincent Poor%
\thanks{The work of R. Liu and H. V. Poor was supported by the
National Science Foundation under Grant CNS-09-05398, and by the Air Force
Office of Scientific Research under Grant FA9550-08-1-0480, and the work of Y.
Liang was supported by a National Science Foundation
CAREER Award under Grant CCF-08-46028 and under Grant CCF-09-15772.}%
\thanks{Ruoheng Liu and H. Vincent Poor are with the Department of Electrical Engineering,
Princeton University, Princeton, NJ 08544, USA  (email: {\{rliu,poor\}@princeton.edu}).}%
\thanks{Yingbin Liang is with the Department of Electrical Engineering and Computer Science, 
Syracuse University, Syracuse, NY 13244, USA (email: yliang06@syr.edu).}%
}


\maketitle



\begin{abstract}
The fading cognitive multiple-access channel with confidential messages (CMAC-CM) is investigated, in which two users attempt to transmit common information to a destination and user 1 also has confidential information intended for the destination. User 1 views user~2 as an eavesdropper and wishes to keep its confidential information as secret as possible from user~2. The multiple-access channel (both the user-to-user channel and the user-to-destination channel) is corrupted by multiplicative fading gain coefficients in addition to additive white Gaussian noise. The channel state information (CSI) is assumed to be known at both the users and the destination. A parallel CMAC-CM with independent subchannels is first studied. The secrecy capacity region of the parallel CMAC-CM is established, which yields the secrecy
capacity region of the parallel CMAC-CM with degraded subchannels. Next, the secrecy capacity region is established for the parallel Gaussian CMAC-CM, which is used to study the fading CMAC-CM. When both users know the CSI, they can dynamically change their transmission powers with the channel realization to
achieve the optimal performance. The closed-form power allocation function that achieves every boundary point of the secrecy capacity region is derived.
\end{abstract}


\section{Introduction}

Wireless transmissions lack physical boundaries and so any adversary within range can receive them. Thus, security is one of the most important issues in wireless communications. One approach to security involves applying encryption algorithms to make messages unintelligible to adversaries. Unfortunately, these
security methods are often designed without consideration of the specific properties of wireless networks. More specifically, encryption methods tend to be layer-specific and ignore the most fundamental communication layer, i.e., the physical-layer, whereby devices communicate through the encoding and
modulation of information into waveforms.

The first study of secure communication via physical layer approaches was captured by a basic wiretap channel introduced by Wyner in \cite{Wyn-BSTJ75}. In this model, a single source-destination communication link is eavesdropped upon by an eavesdropper via a degraded channel. The source node wishes to send
confidential information to the destination node in a reliable manner as well as to keep the eavesdropper as ignorant of this information as possible. The performance measure of interest is the secrecy capacity which characterizes the largest possible communication rate from the source node to the destination
node with the eavesdropper obtaining no source information. Wyner's formulation was generalized by Csisz\'{a}r and K\"{o}rner who determined the secrecy capacity region of a more general model referred to as the broadcast channel with confidential messages (BCC) \cite{CK-IT78}.

More recently, multi-terminal communication with confidential messages has been
studied intensively. (See \cite{LPS-NOW} for a recent survey of progress in
this area.) Among these studies, a generalization of both the wiretap channel
and the classical multiple-access channel (MAC) was studied in
\cite{LP-IT08-MAC}, in which each user also receives channel outputs, and hence
may obtain the confidential information sent by the other user from the channel
output it receives. In this communication scenario, each user views the other
user as an eavesdropper, and wishes to keep its confidential information as
secret as possible from the other user. The authors of \cite{LP-IT08-MAC}
investigated the rate-equivocation region and secrecy capacity region for this
channel. Some other related studies on secure communication over multiple
access channels can be found in \cite{Liu:ISIT:06,Tekin:ISIT:06,SY:CISS:09}.

Fading has traditionally been considered to be an obstacle to providing
reliable wireless communication. However, over the past decade, it has been
demonstrated that fading can help improve capacity, reliability, and
confidentiality of wireless networks. The impact of fading on secure
communication was studied in, e.g.,
\cite{Liang06novit,Gopala:ISIT:07,Tang:IT:09}. More specifically,
\cite{Liang06novit} studied the secrecy capacity of ergodic fading BCCs when
the channel state information (CSI) is known at all communicating nodes;
\cite{Gopala:ISIT:07} considered the ergodic scenario of fading wiretap channel
in which the transmitter has no CSI about the eavesdropper channel; and
\cite{Tang:IT:09} studied the outage preference of secure communication over
wireless channels, in which the transmitter has no CSI about either the
legitimate receiver's channel or the eavesdropper's channel.

In this paper, we investigate the fading cognitive multiple-access channel with both common and confidential messages, a problem which is inspired by the studies of secure communication over MACs in \cite{LP-IT08-MAC}. In our communication scenario, we assume that two users (users 1 and 2) have common information, while user 1 has confidential information intended for a destination and treats
user~2 as an eavesdropper. Hence, user 1 wishes to keep its confidential messages as secret as possible from user 2. We refer to this model as the cognitive MAC with one confidential message (CMAC-CM); (see Fig. 1.(a)), because this channel also models cognitive communication in which the secondary user (user 1) helps the primary user (user 2) to send a common message $W_0$, and also has a confidential message $W_1$ intended for the destination, which needs to be kept secret from the primary user. Furthermore, we consider the situation in which both the user-to-user and the user-to-destination channels are corrupted
by multiplicative fading gain coefficients in addition to additive white Gaussian noise. The fading CMAC-CM model captures the basic time-varying and superposition properties of wireless channels, and thus, understanding this channel plays an important role in solving security issue in wireless application. For the fading CMAC-CM, we assume that the fading gain coefficients are stationary and ergodic over time and that the CSI is known at both users and the destination. Note that knowledge of the user-to-destination CSI is necessary in order to cooperatively transmit the common message, and thus should be provided through state feedback from the destination terminal to the user terminals. Knowledge of CSI between the
user terminals can be obtained via the reciprocity property of those channels. Users are motivated to do so in order to enable better cooperation for sending the common message.

\begin{figure*}
\noindent
\begin{minipage}[b]{0.45\linewidth}
     \centerline{\includegraphics[width=0.85\linewidth,draft=false]{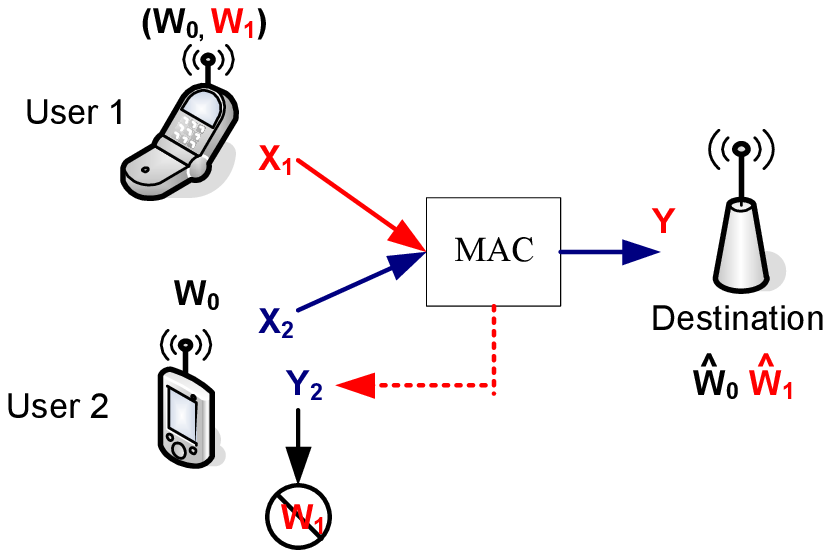}}
     \centerline{\mbox{\footnotesize (a) CMAC-CM}}
  \end{minipage}\hfill
  \begin{minipage}[b]{0.45\linewidth}
     \centerline{\includegraphics[width=0.85\linewidth,draft=false]{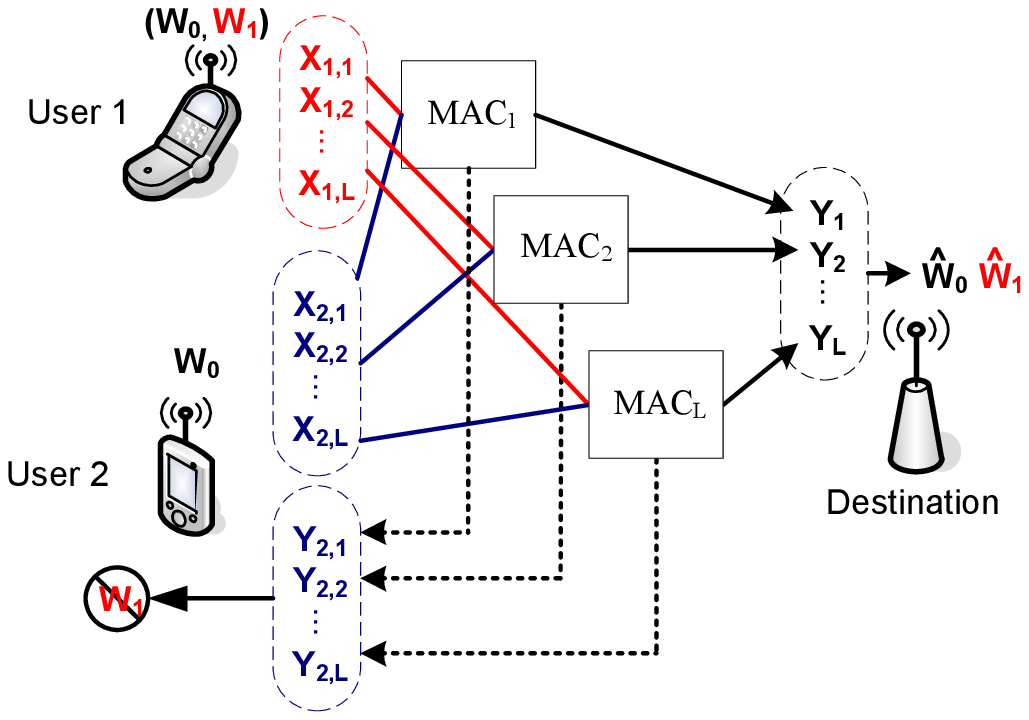}}
    \centerline{\mbox{\footnotesize (b) Parallel CMAC-CM}}
  \end{minipage}
  \caption{Cognitive multiple-access channel with confidential messages.}
  \label{fig:MAC-CM}
\end{figure*}

To solve the fading CMAC-CM problem, we first consider a general information-theoretic model, i.e., the parallel MAC with $L$ independent subchannels. As shown in Fig.~\ref{fig:MAC-CM}.(b), the two users communicate with the destination over $L$ parallel links and each of the $L$ links is eavesdropped upon by user 2. We establish the secrecy capacity region for the parallel CMAC-CM. In particular, we provide a converse proof to show that having independent inputs for each subchannel is optimal to achieve the secrecy capacity region. The secrecy capacity region of the parallel CMAC-CM further gives the secrecy capacity region of the parallel CMAC-CM with degraded subchannels. Next, we consider the parallel Gaussian CMAC-CM, which is an example parallel CMAC-CM with degraded subchannels. Based on the maximum-entropy theorem \cite{Cover} and the extremal inequality \cite{LiuTie:it:07}, we show that the secrecy capacity region of the parallel Gaussian CMAC-CM is achievable by using jointly Gaussian inputs and optimizing power allocations at two users among the parallel subchannels. We then apply this result to investigate the fading CMAC-CM. We study the ergodic performance, where no delay constraint on message transmission is assumed and the secrecy capacity region is averaged over all channel states. In fact, the fading CMAC-CM can be viewed as the parallel Gaussian CMAC-CM with each fading state corresponding to one subchannel. Hence, the secrecy capacity region of the parallel Gaussian CMAC-CM applies to the fading CMAC-CM. Since both users know the CSI, users can dynamically change their transmission powers with the channel realization to achieve the optimal performance. The optimal power allocation that achieves every boundary point of the secrecy capacity region can be characterized as a solution to a non-convex problem. The Karush-Kuhn-Tucker (KKT) conditions (as necessary conditions) greatly facilitate exploitation of the specific structure of the problem, and enable us to obtain a closed-form solution for the optimal power allocation strategy for the two users.

The remainder of this paper is organized as follows. We first study the parallel CMAC-CM with independent subchannels and its special case of the parallel CMAC-CM with degraded subschannels in Section~\ref{sec:PMAC-CM}. Next, we investigate the secrecy capacity region of the parallel Gaussian CMAC-CM in Section~\ref{sec:PGMAC} and the ergodic performance of the fading CMAC-CM in Section~\ref{sec:F-MAC}. We then provide some numerical examples in Section~\ref{sec:ex}. Finally, we summarize our results in Section~\ref{sec:con}.

\section{Parallel CMAC-CM}
\label{sec:PMAC-CM}

\subsection{Channel Model}

We consider the discrete memoryless parallel CMAC-CM with $L$ independent subchannels (see Fig.~\ref{fig:MAC-CM}.(b)). Each subchannel is assumed to connect users 1 and 2 to the destination, and user 2 can also receive the channel output from each subchannel, and hence may obtain information sent by user 1. The channel transition probability distribution is given by
\begin{align}
p(y_{[1,L]},y_{2,[1,L]}&|x_{1,[1,L]},x_{2,[1,L]})
=\prod_{j=1}^{L}p(y_{j},y_{2,j}|x_{1,j},x_{2,j}),
\end{align}
where $y_{[1,L]}:=(y_1,...,y_L)$.

In this model, a common message $W_0$ is known to both the primary user (user 2) and the secondary user (user~1), and hence both users cooperate to transmit $W_0$ to the destination. Moreover, the secondary user (user 1) also has confidential message $W_1$ intended for the destination. User 1 views user 2 as an eavesdropper and wishes to keep its confidential information as secret as possible from user 2. In this paper, we focus on the case in which perfect secrecy is achieved, i.e., user 2 should not obtain any information about the message $W_1$. More formally, this condition is characterized by (e.g., see \cite{Wyn-BSTJ75,CK-IT78,LP-IT08-MAC}):
\begin{align}
\frac{1}{n}I(W_1;Y_{2}^n, X_2^{n},W_0) \rightarrow 0
\end{align}
where $X_2^{n}:=(X_{2,1},\dots,X_{2,n})$ and $Y_{2}^n:=(Y_{2,1},\dots,Y_{2,n})$
are the input and output sequences of user~2, respectively, and the limit is
taken as the block length $n\rightarrow \infty$. The goal is to characterize
the {\it secrecy capacity region} $\Cc_s$ that contains rate pairs achievable
by some coding scheme (more detailed definitions for the rates of the messages and encoding and decoding schemes can be found in \cite{LP-IT08-MAC}).

\subsection{Secrecy Capacity Region of the Parallel CMAC-CM}

For the parallel CMAC-CM, we obtain the following secrecy capacity region.
\begin{theorem} \label{thm:p}
For the parallel CMAC-CM, the secrecy capacity region is given by
\begin{align}
\Cc_s^{\rm [P]}=& \bigcup_{ \substack{\prod_j
p(q_j,x_{2,j})p(u_j|q_j)p(x_{1,j}|u_j)\\p(y_j,y_{2,j}|x_{1,j},x_{2,j})}}
\left\{
\begin{array}{l}
(R_0,\,R_1):\\
~~R_0\ge 0,\; R_1\ge 0;\\
~~R_1\le \sum_{j=1}^{L} [I(U_j; Y_j|X_{2,j},Q_j)
-I(U_j;Y_{2,j}|X_{2,j},Q_j)] \\
~~R_0\le \sum_{j=1}^{L} I(Q_j,X_{2,j}; Y_j)
\end{array}
\right\} \label{eq:p-gb}
\end{align}
where $Q_j$ and $U_j$'s are auxiliary random variables, and $Q_j$ can be chosen
to be a deterministic function of $U_j$ for $j=1,\dots,L$.
\end{theorem}
\begin{proof}
See Appendix~\ref{app:thm1}.
\end{proof}

Theorem 1 implies that having independent inputs for each subchannel is
optimal. This fact does not follow directly from the single-letter result on
the secrecy capacity region of the CMAC-CM given in \cite{LP-IT08-MAC}. Hence, a
converse proof is needed, which is provided in Appendix~\ref{app:thm1}.


\subsection{Parallel CMAC-CM with Degraded Subchannels}

We consider the parallel CMAC-CM with degraded subchannels, in which each
subchannel is either degraded such that given the input of user 2, the output
at user 2 is a conditionally degraded version of the output at the destination,
or reversely degraded such that given the input of user 2, the output at the
destination is a conditionally degraded version of the output at user 2.

Following \cite{LP-IT08-MAC}, we define the conditionally degraded subchannels
as follows. Let $\Ac$ denote the index set that includes all indices of
subchannels such that given $x_{2,j}$, the output at user 2 is a conditionally
degraded version of the output at the destination, i.e., for $j\in \Ac$,
\begin{align}
p(y_j,y_{2,j}|x_{1,j},x_{2,j})=p(y_j|x_{1,j},x_{2,j})p(y_{2,j}|y_{j},x_{2,j}).
\label{eq:def-inda1}
\end{align}
We further define $\bar{\Ac}$ to be the complement of the set $\Ac$, and
$\bar{\Ac}$ includes all indices of subchannels such that given $x_{2,j}$, the
output at the destination is a conditionally degraded version of the output at
user 2, i.e., for $j\in \bar{\Ac}$,
\begin{align}
p(y_j,y_{2,j}|x_{1,j},x_{2,j})=p(y_{2,j}|x_{1,j},x_{2,j})p(y_{j}|y_{2,j},x_{2,j}).
\label{eq:def-inda2}
\end{align}
Hence, the channel transition probability distribution is given by
\begin{align}
p(y_{[1,L]},y_{2,[1,L]}|&x_{1,[1,L]},x_{2,[1,L]})\notag\\
=&\prod_{j\in \Ac} p(y_j|x_{1,j},x_{2,j})p(y_{2,j}|y_{j},x_{2,j})
\prod_{j\in \bar{\Ac}} p(y_{2,j}|x_{1,j},x_{2,j})p(y_{j}|y_{2,j},x_{2,j}).
\label{eq:ch-pd}
\end{align}

For the parallel CMAC-CM with degraded subchannels, we apply Theorem~\ref{thm:p} and
obtain the following secrecy capacity region.

\begin{theorem} \label{thm:pd}
For the parallel CMAC-CM with degraded subchannels, the secrecy capacity region is given
by
\begin{align}
\Cc_s^{\rm [D]}=& \bigcup_{ \substack{ \prod_{j}
p(q_j,x_{2,j})p(x_{1,j}|q_j)\\p(y_j,y_{2,j}|x_{1,j},x_{2,j})}}
\left\{
\begin{array}{l}
(R_0,\,R_1):\\
R_0\ge 0,\; R_1\ge 0;\\
~~R_1\le \sum_{j\in \Ac} [I(X_{1,j}; Y_j|X_{2,j},Q_j)
-I(X_{1,j};Y_{2,j}|X_{2,j},Q_j)] \\
~~R_0\le \sum_{j\in \Ac} I(Q_j,X_{2,j}; Y_j)
+\sum_{j\in \bar{\Ac}} I(X_{1,j},X_{2,j}; Y_j)
\end{array}
\right\}\label{eq:pd-mac}
\end{align}
where $Q_j$, for $j=1,\dots,L$, are auxiliary random variables that satisfy the
Markov chain relationship
\begin{align}
Q_j\rightarrow (X_{1,j},X_{2,j}) \rightarrow (Y_j,Y_{2,j}).
\end{align}
\end{theorem}
\begin{proof}
See Appendix~\ref{app:thm2}.
\end{proof}

It can be seen that the common message $W_0$ is sent
over all subchannels, and the confidential message $W_1$ of user 1 is sent only
over the subchannels for which the output at user 2 is a {\it conditionally
degraded} version of the output at the destination. Furthermore, user 1 sends
the common message $W_0$ and the confidential message $W_1$ by using
superposition encoding.

\section{Parallel Gaussian CMAC-CM} \label{sec:PGMAC}

\subsection{Channel Model}
In this section, we consider the parallel Gaussian CMAC-CM in which the channel outputs at
the destination and user 2 are corrupted by additive Gaussian noise terms. The channel
input-output relationship is given by
\begin{align}
Y_{j,i}&=X_{1,j,i}+X_{2,j,i}+Z_{j,i}\notag\\
\text{and}\qquad Y_{2,j,i}&=X_{1,j,i}+X_{2,j,i}+Z_{2,j,i} \label{eq:pGMAC}
\end{align}
where $i$ is the time index, and for $j=1,\dots,L$, the noise processes
$\{Z_{j,i}\}$ and $\{Z_{2,j,i}\}$ are independent and identically distributed
(i.i.d.) with the components being zero-mean Gaussian random variables with
variances $\nu_j$ and $\mu_j$, respectively. We assume $\nu_j<\mu_j$ for
$j\in\Ac$ and $\nu_j\ge \mu_j$ for $j\in\bar{\Ac}$. The channel input sequences
$X^{n}_{1,[1,L]}$ and $X^{n}_{2,[1,L]}$ are subject to average power
constraints $P_1$ and $P_2$, respectively, i.e.,
\begin{align}
\frac{1}{n}\sum_{i=1}^{n}\sum_{j=1}^{L}\E[X_{1,j,i}^2]&\le
P_1\notag\\
\text{and}\qquad \frac{1}{n}\sum_{i=1}^{n}\sum_{j=1}^{L}\E[X_{2,j,i}^2]&\le
P_2.
\end{align}

\subsection{Secrecy Capacity Region}
We now apply Theorem~2 to obtain the secrecy capacity region of the parallel
Gaussian MAC. It can be seen from (\ref{eq:pGMAC}) that the subchannels of the
parallel Gaussian MAC are not physically degraded. We consider the following
subchannels, for $j\in \Ac$:
\begin{align}
Y_{j,i}&=X_{1,j,i}+X_{2,j,i}+Z_{j,i},~~ Y_{2,j,i}=Y_{j,i}+Z'_{2,j,i};
\label{eq:spGMAC-1}
\end{align}
and, for $j\in \bar{\Ac}$:
\begin{align}
Y_{j,i}&=Y_{2,j,i}+Z'_{j,i}, ~~ Y_{2,j,i}=X_{1,j,i}+X_{2,j,i}+Z_{2,j,i}
\label{eq:spGMAC-2}
\end{align}
where $\{Z'_{j,i}\}$ and $\{Z'_{2,j,i}\}$ are i.i.d. random processes with
components being zero-mean Gaussian random variables with variances
$\nu_j-\mu_j$ for $j\in\bar{\Ac}$ and $\mu_j-\nu_j$ for $j\in\Ac$,
respectively. Moreover, $\{Z'_{j,i}\}$  is independent of $\{Z_{2,j,i}\}$, and
$\{Z'_{2,j,i}\}$  is independent of $\{Z_{j,i}\}$. We notice that the channel
defined in (\ref{eq:spGMAC-1})-(\ref{eq:spGMAC-2}) is a parallel Gaussian MAC
with physically degraded subchannels. Since the channel
(\ref{eq:spGMAC-1})-(\ref{eq:spGMAC-2}) has the same marginal distributions
$p(y|x_1,x_2)$ and $p(y_2|x_1,x_2)$ as the parallel Gaussian MAC defined in
(\ref{eq:pGMAC}), these two channels have the same secrecy capacity
region.\footnote{This argument is in fact identical to the so-called {\it
degraded, same-marginals} technique; e.g., see \cite{LP-IT08-MAC} for further
details.}

For the channel defined in  (\ref{eq:spGMAC-1})-(\ref{eq:spGMAC-2}), we can apply Theorem~\ref{thm:pd} to obtain he following secrecy capacity region. In particular, the degradedness of the subchannels allows the use of the entropy power inequality in the proof of the converse. We can thus obtain the secrecy
capacity region for the parallel Gaussian CMAC-CM.

\begin{theorem} \label{thm:pG-MAC}
For the parallel Gaussian CMAC-CM, the secrecy capacity region is given by
\begin{align}
\Cc_s^{\rm [G]}=& \bigcup_{\underline{p}\in \Pc}
\left\{
\begin{array}{l}
(R_0,\,R_1):\\
~~R_0\ge 0,\; R_1\ge 0;\\
~~R_1\le \sum_{j\in \Ac}
\left[\frac{1}{2}\log\left(1+\frac{b_j}{\nu_j}\right)
-\frac{1}{2}\log\left(1+\frac{b_j}{\mu_j}\right)\right] \\
~~R_0\le \sum_{j\in \Ac}
\frac{1}{2}\log\left(1+\frac{a_j+p_{2,j}+2\sqrt{a_jp_{2,j}} }{b_j+\nu_j}\right)
\\
\qquad + \sum_{j\in \bar{\Ac}} \frac{1}{2}\log\left(1+\frac{a_j+p_{2,j}+2\sqrt{
a_jp_{2,j}}}{\nu_j}\right)
\end{array}
\right\}\label{eq:pG-mac}
\end{align}
where $\underline{p}$ is the power allocation vector, which consists of
$(a_j,b_j,p_{2,j})$ for $j\in\Ac$ and $(a_j,0,p_{2,j})$ for $j\in \bar{\Ac}$ as
components, and the set $\Pc$ includes all power allocation vectors
$\underline{p}$ that satisfy the power constraint
\begin{align}
\Pc:=\left\{\underline{p}: \sum_{j=1}^{L} (a_j+b_j) \le
P_1~\text{and}~\sum_{j=1}^{L} p_{2,j} \le P_2\right\}. \label{eq:pow0}
\end{align}
\end{theorem}
\begin{proof}
See Appendix~\ref{app:thm3}.
\end{proof}


We notice that $\underline{p}$ denotes the power allocation among all
subchannels. In particular, for $j\in \Ac$, since user 1 needs to transmit both
common and confidential information, the pair $(a_j, b_j)$ controls the power
allocation between the common message $W_0$ and the confidential message $W_1$.
For $j\in \bar{\Ac}$, user 1 transmits only the common information, and $b_j=0$
indicates that the power is allocated to transmit the common message $W_0$
only.

\subsection{Optimal Power Allocation}

To characterize the secrecy capacity region of the parallel Gaussian CMAC-CM given in (\ref{eq:pG-mac}), we need to characterize every boundary point and the power allocation vector that achieve each boundary point. Since the secrecy capacity region $\Cc_s^{\rm [G]}$ is convex, for every boundary point $(R_0^{\star},R_1^{\star})$, there exists $\gamma_1 \ge 0$ such that $(R_0^{\star},R_1^{\star})$ is the solution to the optimization problem
\begin{align}
\max_{(R_0,R_1) \in \Cc_s^{\rm [G]}} \left[R_0+ \gamma_1 R_1\right].
\label{eq:opt1}
\end{align}
Note that the optimization problem (\ref{eq:opt1}) serves as a complete
characterization of the corresponding boundary of the secrecy capacity region,
and the solution to (\ref{eq:opt1}) provides the power allocations that achieve
the boundary of the secrecy capacity region. Let $(x)^{+}=\max(0,\,x)$.
We obtain the optimal power allocation $\underline{p}$ that solves (\ref{eq:opt1}).

\begin{theorem} \label{thm:op1}
Let $\underline{p}^{\star}$ be an optimal solution to the optimization problem
of (\ref{eq:opt1}) that achieves the boundary of the secrecy capacity region of
the parallel Gaussian CMAC-CM. Then, $\underline{p}^{\star}$ can be written as follows.

For $j\in \Ac$, if
\begin{align}
\frac{2\lambda_1^2\ln2}{\lambda_1+\lambda_2} <
\frac{\gamma_1(\mu_j-\nu_j)-\mu_j}{\mu_j\nu_j},
\end{align}
then
\begin{align}
a_j^{\star}&
=\frac{\lambda_2^2}{(\lambda_1+\lambda_2)^2}\left(s_{1,j}-\phi_j\right)^{+}, \notag\\
b_j^{\star}&=\left( \min\left[s_{2,j},\,\phi_j \right]\right)^{+} \notag\\
\text{and}\qquad p_{2,j}^{\star}&=\frac{\lambda_1^2}{(\lambda_1+\lambda_2)^2}\left(s_{1,j}-\phi_j\right)^{+};
\end{align}
alternatively, if
\begin{align}
\frac{2\lambda_1^2\ln2}{\lambda_1+\lambda_2} \ge
\frac{\gamma_1(\mu_j-\nu_j)-\mu_j}{\mu_j\nu_j},
\end{align}
then
\begin{align}
a_j^{\star}&=\frac{\lambda_2^2}{(\lambda_1+\lambda_2)^2}\left(s_{1,j}\right)^{+}, \notag\\
b_j^{\star}&=0 \notag\\
\text{and} \qquad
p_{2,j}^{\star}&=\frac{\lambda_1^2}{(\lambda_1+\lambda_2)^2}\left(s_{1,j}\right)^{+};
\end{align}
for $j\in \bar{\Ac}$,
\begin{align}
a_j^{\star}=\frac{\lambda_2^2}{(\lambda_1+\lambda_2)^2}\left(s_{1,j}\right)^{+}
\quad \text{and} \quad
p_{2,j}^{\star}&=\frac{\lambda_1^2}{(\lambda_1+\lambda_2)^2}\left(s_{1,j}\right)^{+};
\end{align}
where $\gamma_1 \ge 0$,
\begin{align}
s_{1,j}&=\frac{\lambda_1+\lambda_2}{2\lambda_1\lambda_2\ln 2}-\nu_j , \notag\\
s_{2,j}&=\frac{1}{2}\left[\sqrt{(\mu_j-\nu_j)\left(\mu_j-\nu_j+\frac{2\gamma_1}{\lambda_1\ln
2}\right)}-(\mu_j+\nu_j)\right], \notag\\
\phi_j&=-\frac{1}{2}\left(\mu_j+\nu_j+\frac{1}{\omega}\right)+\frac{1}{2}\sqrt{\left(\mu_j+\nu_j+
\frac{1}{\omega}\right)^2-
4\left[\mu_j\nu_j-\frac{\gamma_1(\mu_j-\nu_j)-\mu_j}{\omega}\right]},\notag\\
\omega&=(2\ln2)\frac{\lambda_1^2}{\lambda_1+\lambda_2}
\end{align}
and the pair $(\lambda_1,\lambda_2)$ is chosen to satisfy the power constraint
\begin{align}
\sum_{j=1}^{L} (a_j+b_j) \le P_1~\text{and}~\sum_{j=1}^{L} p_{2,j} \le P_2.
\label{eq:pow1}
\end{align}
\end{theorem}
\begin{proof}
The optimization problem is non-convex. Our proof technique involves applying KKT conditions (as necessary conditions), which help express the Lagrangian in the form of an integral. This specific structure of the problem is then exploited to obtain a closed-form solution for the optimal power allocation strategy. The details can be found in Appendix~\ref{app:op}.
\end{proof}


\section{Fading CMAC-CM}  \label{sec:F-MAC}

In this section, we study the fading CMAC-CM, where both the user-to-destination and the
user-to-user channels are corrupted by multiplicative fading gain processes in addition
to additive white Gaussian processes. The channel input-output relationship is given by
\begin{align}
Y_i&=h_{1,i}X_{1,i}+h_{2,i}X_{2,i}+Z_i \notag\\
\text{and} \quad Y_{2,i}&=g_{1,i}X_{1,i}+g_{2,i}X_{2,i}+Z_{2,i}
\end{align}
where $i$ is the time index, $X_{1,i}$ and $X_{2,i}$ are channel inputs at the
time instant $i$ from user 1 and user 2, respectively, $Y_i$ and $Y_{2,i}$ are
channel outputs at the time instant $i$ at the destination and the receiver of
user 2, respectively; $\hu_i:=(h_{1,i},h_{2,i})$ and $\gu_i:=(g_{1,i},g_{2,i})$
are proper complex random channel attenuation pairs imposed on the destination
and the receiver of user 2; and the noise processes $\{Z_i\}$ and $\{Z_{2,i}\}$
are i.i.d. with the components being zero-mean proper complex Gaussian random
variables with variances $\nu$ and $\mu$, respectively. The input sequences
$\{X_{1,i}\}$ and $\{X_{2,i}\}$ are subject to the average power constraint
$P_1$ and $P_2$, i.e.,
\begin{align}
\frac{1}{n}\sum_{i=1}^{n}\E[X_{1,i}^2]\le P_1 \quad \text{and} \quad
\frac{1}{n}\sum_{i=1}^{n}\E[X_{2,i}^2]\le P_2.
\end{align}
\begin{figure}[bt]
 \centerline{\includegraphics[width=0.5\linewidth,draft=false]{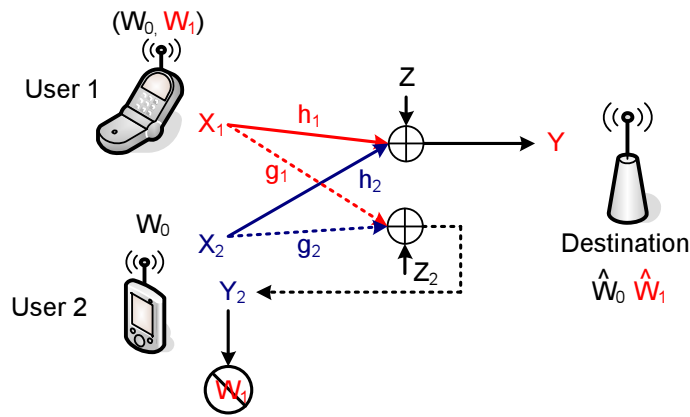}}
  \caption{Fading CMAC-CM.}
  \label{fig:fMAC}
\end{figure}

We assume that the CSI (i.e., the realization of $(\hu,\gu)$) is known at both
the transmitters and the receivers instantaneously. Depending on the CSI, each
user can dynamically change its transmission power and rate to achieve better
performance. In this section, we assume that there is no delay constraint on
the transmitted messages, and that the secrecy capacity region is an average
over all channel states, which is referred to as the {\it ergodic} secrecy
capacity region.

We notice that for a given fading state, i.e., a realization of $(\hu,\gu)$, the fading
CMAC-CM is a Gaussian CMAC-CM. Hence, the fading CMAC-CM can be viewed as a parallel
Gaussian CMAC-CM with each fading state corresponding to one subchannel. Thus, the
following secrecy capacity region of the fading CMAC-CM follows from
Theorem~\ref{thm:pG-MAC}.

In the following, for each channel state $(\hu,\gu)$, we use $p_1(\hu,\gu)$ and
$p_2(\hu,\gu)$ to denote the powers allocated at users 1 and 2, respectively.
We further define
\begin{align}
p(\hu,\gu):= \left(a(\hu,\gu), b(\hu,\gu), p_2(\hu,\gu)\right).
\end{align}
Let $\Pc$ denote the set that includes all power allocations that satisfy the
power constraint
\begin{align}
\Pc:=\bigl\{p(\hu,\gu):~ &\E[a(\hu,\gu)+b(\hu,\gu)] \le P_1
\quad \text{and} \quad \E[p_2(\hu,\gu)] \le P_2 \bigr\}, \label{eq:pow2}
\end{align}
and $\Ac$ denote the set of channel states as follows:
\begin{align}
\Ac:=\left\{ (\hu,\gu):~ \frac{|h_1|^2}{\nu}>\frac{|g_1|^2}{\mu}\right\}.
\end{align}

\begin{corollary} \label{cor:f-MAC}
The secrecy capacity region of the fading CMAC-CM is given by (\ref{eq:F-mac})
\begin{align}
\Cc_s^{\rm [F]}= &\bigcup_{p(\hu,\gu) \in \Pc}
\left\{
\begin{array}{l}
(R_0,\,R_1):\\
~~R_0\ge 0,\; R_1\ge 0;\\
~~R_1\le \E_{(\hu,\gu) \in \Ac} \left[\log\left(1+
\frac{b(\hu,\gu)|h_1|^2}{\nu}\right)
-\log\left(1+\frac{b(\hu,\gu)|g_1|^2}
{\mu}\right)\right] \\
~~R_0\le \E_{(\hu,\gu) \in \Ac} \log\left(1+\frac{\chi(\hu,\gu)}
{b(\hu,\gu)|h_1|^2+\nu}\right)
+ \E_{(\hu,\gu)\in \bar{\Ac}}
\log\left(1+\frac{\chi(\hu,\gu)}{\nu}\right)
\end{array}
\right\}\label{eq:F-mac}
\end{align}
where
\begin{align}
\chi(\hu,\gu)&=\left[\sqrt{a(\hu,\gu)}|h_1| +\sqrt{p_2(\hu,\gu)}|h_2|\right]^2
\end{align}
and the random vector pair $(\hu,\gu)$ has the same distribution as the
marginal distribution of the process $\{(\hu_i,\gu_i)\}$ at a single time
instant.
\end{corollary}

The secrecy capacity region given in Corollary~\ref{cor:f-MAC} is established for fading
processes $(\hu,\gu)$ where only ergodic and stationary conditions are assumed. The
fading process $(\hu,\gu)$ can be correlated across time, and is not necessarily
Gaussian.

Since users are assumed to know the CSI, they can allocate their powers according to the
instantaneous channel realization to achieve the optimal performance, i.e., the boundary of
the secrecy capacity region. The optimal power allocation that achieves the boundary of
the secrecy capacity region for the fading CMAC-CM can be derived from
Theorem~\ref{thm:op1} and is given in the following.

\begin{corollary} \label{cor:op2}
Let $p(\hu,\gu)^{\star}$ be an optimal power allocation that achieves the boundary of the
secrecy capacity region of the fading CMAC-CM. Then, $p(\hu,\gu)^{\star}$ is given as follows:
\begin{itemize}
  \item for $(\hu,\gu)\in \Ac$, if
  \begin{align}
   \frac{\lambda_1^2|h_2|^2\ln2}{\lambda_1|h_2|^2+\lambda_2|h_1|^2} <
    \frac{\gamma_1\left(\mu|h_1|^2-\nu|g_1|^2\right)-\mu|h_1|^2}{\mu\nu},
  \end{align}
  then
  \begin{align}
    a^{\star}(\hu,\gu)&=\frac{\lambda_2^2|h_1|^2}{\left(\lambda_1|h_2|^2+\lambda_2|h_1|^2\right)^2}\left[s_1(\hu,\gu)-\phi(\hu,\gu)\right]^{+},\notag\\
    b^{\star}(\hu,\gu)&=\left(\min\left[s_2(\hu,\gu),\,\phi(\hu,\gu)\right]\right)^{+}\notag\\
    \text{and} \qquad
    p_2^{\star}(\hu,\gu)&=\frac{\lambda_1^2|h_2|^2}{\left(\lambda_1|h_2|^2+\lambda_2|h_1|^2\right)^2}\left[s_1(\hu,\gu)-\phi(\hu,\gu)\right]^{+};
  \end{align}
  alternatively, if
  \begin{align}
   \frac{\lambda_1^2|h_2|^2\ln2}{\lambda_1|h_2|^2+\lambda_2|h_1|^2} \ge
    \frac{\gamma_1\left(\mu|h_1|^2-\nu|g_1|^2\right)-\mu|h_1|^2}{\mu\nu},
  \end{align}
  then
  \begin{align}
    a^{\star}(\hu,\gu)&=\frac{\lambda_2^2|h_1|^2}{\left(\lambda_1|h_2|^2+\lambda_2|h_1|^2\right)^2}\left[s_1(\hu,\gu)\right]^{+},\notag\\
    b^{\star}(\hu,\gu)&=0\notag\\
    \text{and} \qquad
    p_2^{\star}(\hu,\gu)&=\frac{\lambda_1^2|h_2|^2}{\left(\lambda_1|h_2|^2+\lambda_2|h_1|^2\right)^2}\left[s_1(\hu,\gu)\right]^{+};
  \end{align}

\item for $(\hu,\gu)\in \bar{\Ac}$,
  \begin{align}
    a^{\star}(\hu,\gu)&=\frac{\lambda_2^2|h_1|^2}{\left(\lambda_1|h_2|^2+\lambda_2|h_1|^2\right)^2}\left[s_1(\hu,\gu)\right]^{+}\notag\\
    \text{and} \qquad
    p_2^{\star}(\hu,\gu)&=\frac{\lambda_1^2|h_2|^2}{\left(\lambda_1|h_2|^2+\lambda_2|h_1|^2\right)^2}\left[s_1(\hu,\gu)\right]^{+};
  \end{align}
\end{itemize}
where $\gamma_1 \ge 0$,
\begin{align}
s_1(\hu,\gu)&=\frac{\lambda_1|h_2|^2+\lambda_2|h_1|^2}{\lambda_1\lambda_2\ln 2}-\nu , \notag\\
s_2(\hu,\gu)&=\frac{1}{2}\left[\sqrt{\left(\frac{\mu}{|g_1|^2}-\frac{\nu}{|h_1|^2}\right)
               \left(\frac{\mu}{|g_1|^2}-\frac{\nu}{|h_1|^2}+\frac{2\gamma_1}{\lambda_1\ln 2}\right)}
               -\left(\frac{\mu}{|g_1|^2}+\frac{\nu}{|h_1|^2}\right)\right], \notag\\
\phi(\hu,\gu)&=-\frac{1}{2}\left(\frac{\mu}{|g_1|^2}+\frac{\nu}{|h_1|^2}+\frac{1}{\omega(\hu,\gu)}\right)\notag\\
  &\quad+\frac{1}{2}\sqrt{\left(\frac{\mu}{|g_1|^2}+\frac{\nu}{|h_1|^2}+
\frac{1}{\omega(\hu,\gu)}\right)^2-
4\left[\frac{\mu}{|g_1|^2}\frac{\nu}{|h_1|^2}-\frac{\gamma_1\left(\frac{\mu}{|g_1|^2}-\frac{\nu}{|h_1|^2}\right)-\frac{\mu}{|g_1|^2}}{\omega(\hu,\gu)}\right]} \notag\\
\omega(\hu,\gu)&=(\ln2)\frac{\lambda_1^2|h_2|^2}{\lambda_1|h_2|^2+\lambda_2|h_1|^2}
\end{align}
and the pair $(\lambda_1,\lambda_2)$ is chosen to satisfy the power constraint
\begin{align}
\E[a(\hu,\gu)+b(\hu,\gu)] \le P_1
\quad \text{and} \quad \E[p_2(\hu,\gu)] \le P_2. \label{eq:pow3}
\end{align}
\end{corollary}

\section{Numerical Examples} \label{sec:ex}

In this section, we study two numerical examples to illustrate the secrecy
capacity regions of the parallel Gaussian CMAC-CM and the fading CMAC-CM,
respectively.

\psfrag{A}{$\Rc_s^{\rm [G]}$}

\psfrag{B}{$\Cc_s^{\rm [G]}$}

\psfrag{C}{coherent combining gain}

\begin{figure}[t]
 \centerline{\includegraphics[width=0.7\linewidth,draft=false]{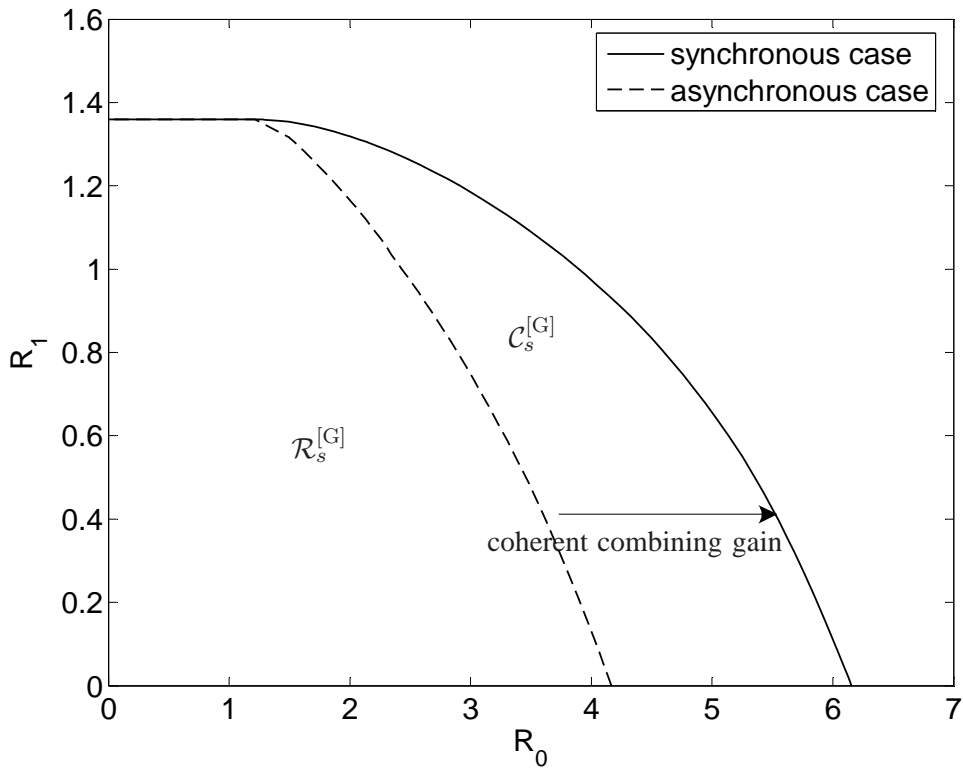}}
\caption{Secrecy capacity region vs. asynchronous secrecy rate region for the
example $L=10$ parallel Gaussian CMAC-CM.}
  \label{fig:sim1}
\end{figure}

We first consider an $L=10$ parallel Gaussian CMAC-CM. We assume that the source
power constraints of users 1 and 2 are
\[P_1=12~\text{dB} \quad \text{and} \quad P_2=10~\text{dB},\]
and the noise variances at the receivers of the destination and of user 2 are
given by
\begin{align*}
\underline{\nu}&=[1, 2, 3, 4, 5, 6, 7, 8, 9, 10] \\
\text{and}\qquad \underline{\mu}&=[5, 3, 4, 9, 1, 10, 8, 7, 2, 6].
\end{align*}
Fig.~\ref{fig:sim1} illustrates the boundary of the secrecy capacity region for
this channel. For comparison, we also consider the asynchronous case, in which
users 1 and 2 send the common message $W_0$ in a asynchronous transmission
mode. In this case, the secrecy rate region is given by
\begin{align}
\Rc_s^{\rm [G]}=& \bigcup_{\underline{p}\in \Pc}
\left\{
\begin{array}{l}
(R_0,\,R_1):\\
~~R_0\ge 0,\; R_1\ge 0;\\
~~R_1\le \sum_{j\in \Ac}
\left[\frac{1}{2}\log\left(1+\frac{b_j}{\nu_j}\right)
\frac{1}{2}\log\left(1+\frac{b_j}{\mu_j}\right)\right] \\
~~R_0\le \sum_{j\in \Ac}
\frac{1}{2}\log\left(1+\frac{a_j+p_{2,j}}{b_j+\nu_j}\right)
+ \sum_{j\in \bar{\Ac}}
\frac{1}{2}\log\left(1+\frac{a_j+p_{2,j}}{\nu_j}\right)
\end{array}
\right\}\label{eq:ApG-mac}
\end{align}
where $\underline{p}$ is the power allocation vector, which consists of
$(a_j,b_j,p_{2,j})$ for $j\in\Ac$ and $(a_j,0,p_{2,j})$ for $j\in \bar{\Ac}$ as
components, and the set $\Pc$ includes all power allocation vector
$\underline{p}$ that satisfy the power constraint (\ref{eq:pow1}). We observe
that the synchronous transmission mode significantly increases the rate $R_0$
of the common message since coherent combining detection can be employed at the
destination.

\begin{figure}[t]
 \centerline{\includegraphics[width=0.7\linewidth,draft=false]{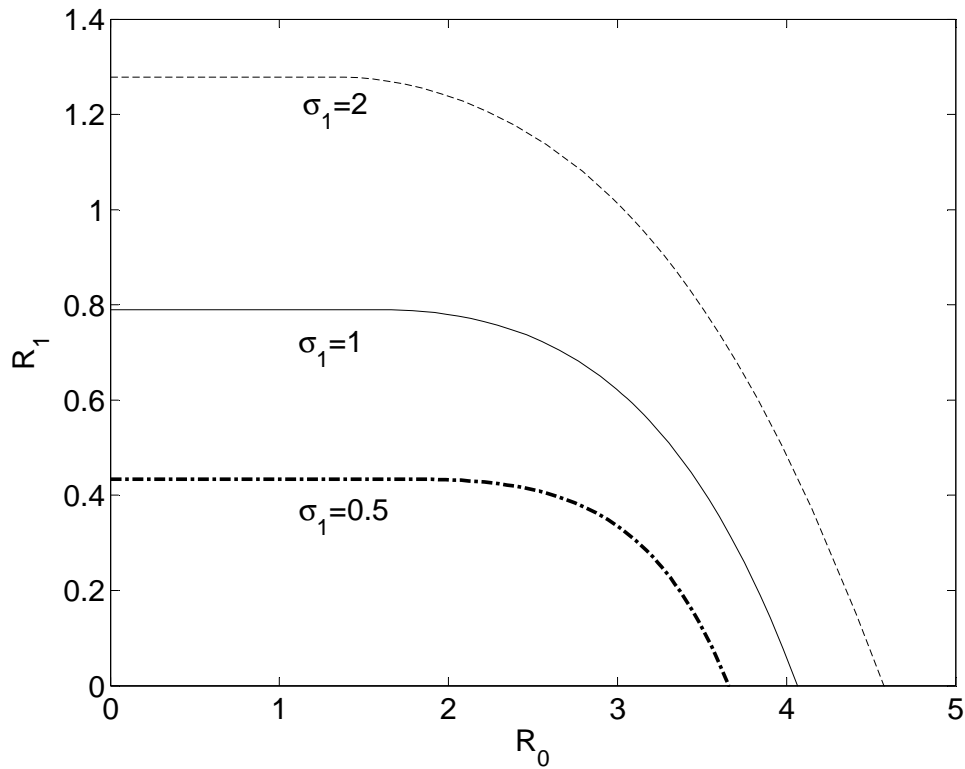}}
\caption{Secrecy capacity regions for the example fading CMAC-CMs
($P_1=P_2=10~\text{dB}$, $\nu=\mu=2$, and $\sigma_2=\sigma_3=1$).}
  \label{fig:sim2}
\end{figure}

\begin{figure}[t]
 \centerline{\includegraphics[width=0.7\linewidth,draft=false]{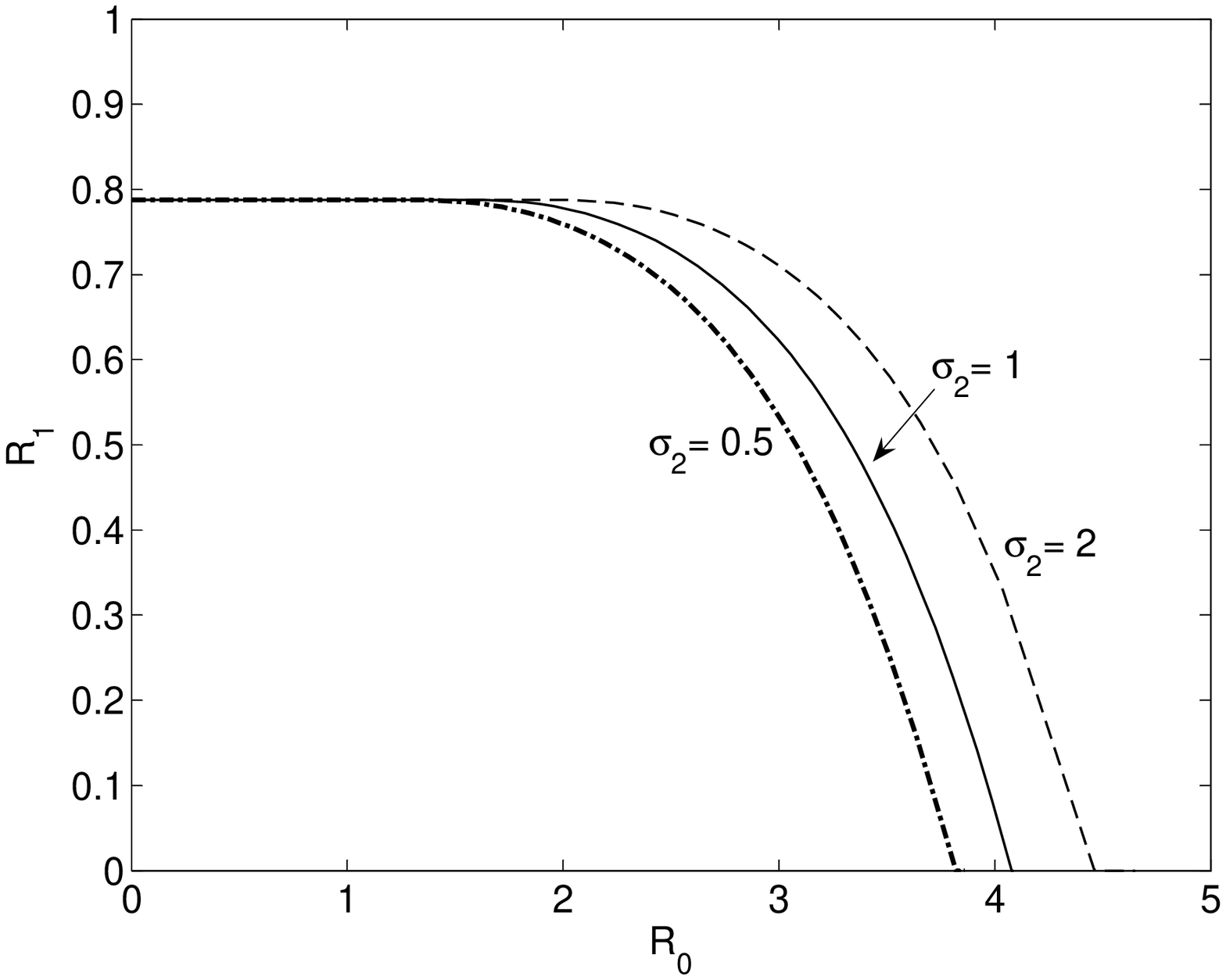}}
\caption{Secrecy capacity regions for the example fading CMAC-CMs
($P_1=P_2=10~\text{dB}$, $\nu=\mu=2$, and $\sigma_1=\sigma_3=1$).}
  \label{fig:sim3}
\end{figure}

\begin{figure}[t]
 \centerline{\includegraphics[width=0.7\linewidth,draft=false]{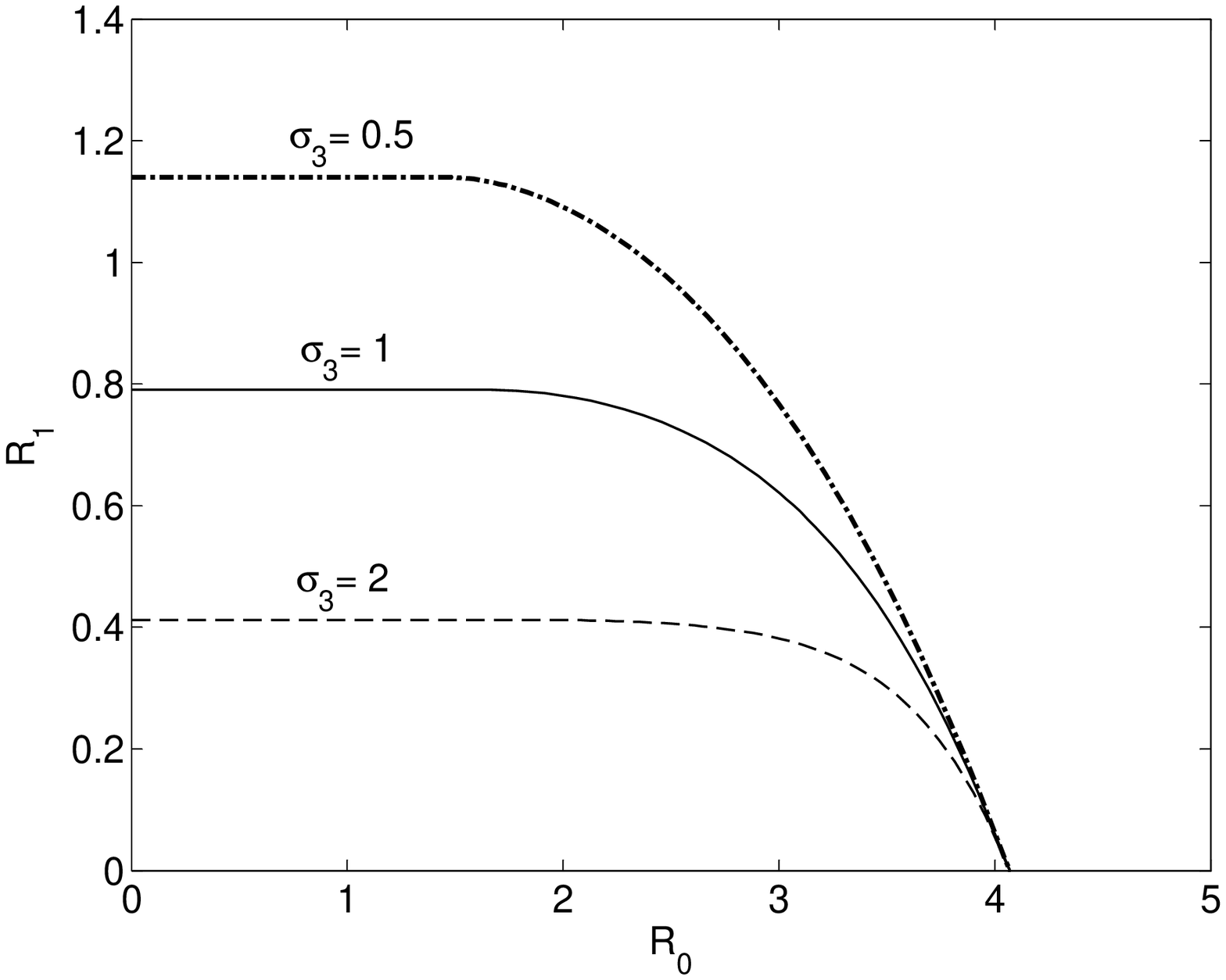}}
\caption{Secrecy capacity regions for the example fading CMAC-CMs
($P_1=P_2=10~\text{dB}$, $\nu=\mu=2$, and $\sigma_1=\sigma_2=1$).}
  \label{fig:sim4}
\end{figure}

Next, we consider the Rayleigh-fading CMAC-CM, where $h_1$, $h_2$ and $g_1$ are
zero-mean proper complex Gaussian random variables. Hence, $|h_1|^2$, $|h_2|^2$
and $|g_1|^2$ are exponentially distributed with means $\sigma_1$, $\sigma_2$
and $\sigma_3$.
We assume that the power constraints of users 1 and 2 are
$P_1=P_2=10~\text{dB}$, and the noise variances at the receivers of the
destination and of user 2 are $\nu=\mu=2$.  In Fig.~\ref{fig:sim2}, we plot the
boundaries of the secrecy capacity regions corresponding to
$\sigma_1=0.5,\;1,\;2$ and fixed $\sigma_2=\sigma_3=1$. It can been seen that as
$\sigma_1$ increases, both the secrecy rate $R_1$ of the confidential message
$W_1$ and the rate $R_0$ of the common message $W_0$ improve. This is because
larger $\sigma_1$ implies a better channel from user 1 to the destination.
In Fig.~\ref{fig:sim3}, we plot the boundaries of the secrecy capacity regions corresponding to
$\sigma_2=0.5,\;1,\;2$ and fixed $\sigma_1=\sigma_3=1$. It can been seen that as
$\sigma_2$ increases, only the rate $R_0$ of the common message $W_0$ improves.
In Fig.~\ref{fig:sim4}, we plot the boundaries of the secrecy capacity regions corresponding to
$\sigma_3=0.5,\;1,\;2$ and fixed $\sigma_1=\sigma_2=1$. It can been seen that as
$\sigma_3$ decreases, only the rate $R_1$ of the confidential message $W_1$ improves.

\section{Conclusion}  \label{sec:con}

We have established the secrecy capacity region of the parallel CMAC-CM, in
which it is seen that having independent inputs to each subchannel is optimal.
From this result, we have derived the secrecy capacity region for the parallel
Gaussian CMAC-CM and the ergodic secrecy capacity region for the fading CMAC-CM.
We have illustrated that, when both users know the CSI, they can
dynamically adapt their transmission powers with the channel realization to
achieve the optimal performance.

  \appendix

\subsection{Proof of Theorem~\ref{thm:p}} \label{app:thm1}

\subsubsection*{Achievability}


The achievability follows from \cite[Corollary~3]{LP-IT08-MAC} by setting
\begin{align}
Q&:=(Q_1,\dots,Q_L), \quad U:=(U_1,\dots,U_L)\notag\\
X_1&:=(X_{1,1},\dots,X_{1,L}), \quad X_2:=(X_{2,1},\dots,X_{2,L})\notag\\
Y&:=(Y_{1},\dots,Y_{L}), \quad \text{and} \quad Y_2:=(Y_{2,1},\dots,Y_{2,L})
\end{align}
with $Q$, $U$, $X_1$, and $X_2$ having independent components. Furthermore, we
choose the components of these random vectors to satisfy the condition
\begin{align}
p(q_j,u_j,x_{1,j},x_{2,j},y_{j},y_{2,j})=p(q_j,x_{2,j})p(u_j|q_j)p(x_{1,j}|u_j)p(y_{j},y_{2,j}|x_{1,j},x_{2,j})
~~\text{for}~j=1,\dots,L.
\end{align}
Using the above definition, we have the following achievable region
\begin{align}
\Rc_s^{P}:=\bigcup_{ \substack{\prod_j
p(q_j,x_{2,j})p(u_j|q_j)p(x_{1,j}|u_j)\\p(y_j,y_{2,j}|x_{1,j},x_{2,j})}}\left\{
(R_0,\,R_1)\;\left|
\begin{array}{l}
R_0\ge 0,\; R_1\ge 0;\\
R_1\le \sum_{j=1}^{L} \left[I(U_j; Y_j|X_{2,j},Q_j)-I(U_j;Y_{2,j}|X_{2,j},Q_j)\right] \\
R_0+R_1\le \sum_{j=1}^{L} \left[ I(U_j,X_{2,j},Q_j; Y_j)
-I(U_j;Y_{2,j}|X_{2,j},Q_j)\right]
\end{array}
\right.\right\}.
\end{align}
Note that
\begin{align}
\left[I(U_j; Y_j|X_{2,j},Q_j)-I(U_j;Y_{2,j}|X_{2,j},Q_j)\right]+I(X_{2,j},Q_j;
Y_j)= I(U_j,X_{2,j},Q_j; Y_j) -I(U_j;Y_{2,j}|X_{2,j},Q_j)
\end{align}
and hence, any rate pair $(R_0,R_1)\in \Cc_s^{P}$ must also satisfies
$(R_0,R_1)\in \Rc_s^{P}$. This implies that the secrecy rate region $\Cc_s^{\rm
[P]}$ is achievable.

\subsubsection*{Converse}

By Fano's inequality \cite[Chapter~2.11]{Cover}, we have
\begin{align}
H\left(W_0,W_1|Y^{n}_{[1,L]}\right) &\le n(R_0,+R_1)\epsilon+1:=n\delta
\label{eq:Fano1}
\end{align}
where $\delta \rightarrow 0$ if $\epsilon \rightarrow 0$. On the other hand,
the information theoretic secrecy implies that
\begin{align}
H(W_1)\le H\left(W_1|Y^{n}_{2,[1,L]},X^{n}_{2,[1,L]}, W_0 \right)+n\epsilon.
\label{eq:s-rq}
\end{align}

Now, we consider the upper bound on the secrecy rate $R_1$ as
\begin{align}
nR_1 &=H(W_1)\notag\\
&\le H\left(W_1|Y^{n}_{2,[1,L]}, X^{n}_{2,[1,L]},W_0\right)+n\epsilon \label{eq:drb-1}\\
&\le H\left(W_1|Y^{n}_{2,[1,L]}, X^{n}_{2,[1,L]},W_0\right)
   - H\left(W_1|Y^{n}_{[1,L]}, X^{n}_{2,[1,L]}, W_0 \right)+n(\epsilon+\delta) \label{eq:drb-2}\\
&= I\left(W_1;Y^{n}_{[1,L]}| X^{n}_{2,[1,L]}, W_0 \right)-I\left(W_1;
Y^{n}_{2,[1,L]}| X^{n}_{2,[1,L]}, W_0\right)+n(\epsilon+\delta) \notag\\
&= \sum_{j=1}^{L}\left[ I\left(W_1;Y^{n}_{j}|Y^{n}_{[1,j-1]}, X^{n}_{2,[1,L]},
W_0\right) - I\left(W_1; Y^{n}_{2,j} |Y^{n}_{2,[j+1,L]},
X^{n}_{2,[1,L]}, W_0\right)\right]+n(\epsilon+\delta)  \label{eq:drb-3}\\
&= \sum_{j=1}^{L}\sum_{i=1}^{n}\left[ I\left(W_1;Y_{j,i}|Y_j^{i-1},
Y^{n}_{[1,j-1]}, X^{n}_{2,[1,L]}, W_0\right) - I\left(W_1; Y_{2,j,i}
|Y^{n}_{2,j,i+1},Y^{n}_{2,[j+1,L]}, X^{n}_{2,[1,L]},
W_0\right)\right]+n(\epsilon+\delta)  \label{eq:drb-4}
\end{align}
where \eqref{eq:drb-1} follows from the secrecy constraint \eqref{eq:s-rq},
\eqref{eq:drb-2} follows from Fano's inequality \eqref{eq:Fano1}, and
\eqref{eq:drb-3} and \eqref{eq:drb-4} follow from the chain rule of mutual
information \cite[Chapter~2.5]{Cover}. Let
\begin{align}
Q_{j,i}:=\left(Y_j^{i-1}, Y^{n}_{[1,j-1]}, Y^{n}_{2,j,i+1},Y^{n}_{2,[j+1,L]},
X^{n}_{2,[1,L]}, W_0 \right). \label{eq:def-Qij}
\end{align}
We notice that this definition implies the Markov chain relationship
\begin{align}
X_{2,j,i}\rightarrow Q_{j,i} \rightarrow W_1 \rightarrow X_{1,j,i}.
\end{align}

Then, following from \cite[Lemma~7]{CK-IT78}, we have
\begin{align}
nR_1&\le \sum_{j=1}^{L}\sum_{i=1}^{n}\left[
I\left(W_1;Y_{j,i}|X_{2,j,i},Q_{j,i}\right) - I\left(W_1; Y_{2,j,i} |X_{2,j,i},
Q_{j,i} \right)\right]+n(\epsilon+\delta). \label{eq:drb-5}
\end{align}

We also can write
\begin{align}
nR_0&=H(W_0)\notag\\
&\le I(W_0;Y^{n}_{[1,L]})+n\delta\label{eq:ss-1}\\
&= \sum_{j=1}^{L}\sum_{i=1}^{n}I(W_0;Y_{j,i}|Y_j^{i-1},
Y^{n}_{[1,j-1]})+n\delta\label{eq:ss-2}\\
&\le \sum_{j=1}^{L}\sum_{i=1}^{n}I(W_0, Y_j^{i-1}, Y^{n}_{[1,j-1]},
Y^{n}_{2,j,i+1},Y^{n}_{2,[j+1,L]}, X^{n}_{2,[1,L]};
Y_{j,i})+n\delta\notag\\
&=\sum_{j=1}^{L}\sum_{i=1}^{n}I(Q_{j,i},X_{2,j,i}; Y_{j,i})+n\delta
\label{eq:ss-4}
\end{align}
where \eqref{eq:ss-1} follows from Fano's inequality \eqref{eq:Fano1},
\eqref{eq:ss-2} follows from the chain rule, and \eqref{eq:ss-4} follows from
the definition of $Q_{j,i}$ in \eqref{eq:def-Qij}.

We introduce a time-sharing random variable $T$ \cite[Chapter~14.3]{Cover} that
is independent of all other random variables in the model, and uniformly distributed over
$\{1,\dots,n\}$. Define $Q_j=(T,Q_i,j)$, $U_j=(Q_j,W_1)$, $X_{1,j}=X_{1,T,j}$,
$X_{2,j}=X_{2,T,j}$, $Y_{2,j}=X_{2,T,j}$, and $Y_{j}=Y_{T,j}$ for
$j=1,\dots,L$. Note that $(Q_j,X_{1,j},X_{2,j},Y_{j},Y_{2,j})$ satisfies the
following Markov chain relationship
\begin{align}
Q_{j}\rightarrow U_j \rightarrow (X_{1,j},X_{2,j}) \rightarrow (Y_j,Y_{2,j}),
\quad \text{for}~j=1,\dots,L.
\end{align}
Using the above definition, \eqref{eq:drb-5} and \eqref{eq:ss-4} become
\begin{align}
R_1 &\le \sum_{j=1}^{L}\left[ I\left(U_j;Y_{j}|X_{2,j},Q_{j}\right) -
I\left(U_j; Y_{2,j} |X_{2,j}, Q_{j} \right)\right]+(\epsilon+\delta).\notag\\
\text{and}\qquad R_0 &\le \sum_{j=1}^{L} I\left(X_{2,j}, Q_{j}
;Y_{j}\right)+\delta.
\end{align}

\subsection{Proof of Theorem~\ref{thm:pd}} \label{app:thm2}
The achievability follows from Theorem~\ref{thm:p} by setting
\begin{align}
U_{j}&=X_{1,j} \qquad \text{for} \quad j \in \Ac\notag\\
\text{and}\qquad Q_{j}=U_{j}&=X_{1,j} \qquad \text{for} \quad j \in \bar{\Ac}.
\end{align}

To show the converse, we first consider the upper bound on $R_0$. By using
\eqref{eq:p-gb} in Theorem~\ref{thm:p}, we have
\begin{align}
R_0 &\le \sum_{j=1}^{L} I(Q_j,X_{2,j}; Y_j)\notag\\
&=\sum_{j\in \Ac} I(Q_j,X_{2,j}; Y_j)+\sum_{j\in \bar{\Ac}} I(Q_{j},X_{2,j};
Y_j)\notag\\
&\le \sum_{j\in \Ac} I(Q_j,X_{2,j}; Y_j)+\sum_{j\in \bar{\Ac}}
I(X_{1,j},X_{2,j}; Y_j) \label{eq:ap-dp1}
\end{align}
where \eqref{eq:ap-dp1} follows from the Markov chain relationships
\begin{align}
&Q_j\rightarrow (X_{1,j},X_{2,j})\rightarrow Y_{j}.
\end{align}

Now, we consider the upper bound on $R_1$. By applying Theorem~\ref{thm:p}, we
obtain
\begin{align}
R_1 &\le \sum_{j=1}^{L} [I(U_j; Y_j|X_{2,j},Q_j)-I(U_j; Y_{2,j}|X_{2,j},Q_j)]\notag\\
&=\sum_{j\in \Ac} [I(U_j; Y_j|X_{2,j},Q_j)-I(U_j; Y_{2,j}|X_{2,j},Q_j)]+
\sum_{j\in \bar{\Ac}} [I(U_j; Y_j|X_{2,j},Q_j)-I(U_j; Y_{2,j}|X_{2,j},Q_j)].
\label{eq:ap-dp2}
\end{align}
For $j\in \bar{\Av}$, the subchannel satisfies
\begin{align}
p(y_j,y_{2,j}|x_{1,j},x_{2,j})=p(y_{2,j}|x_{1,j},x_{2,j})p(y_{j}|y_{2,j},x_{2,j}),
\quad \text{for}~j\in \bar{\Ac}.
\end{align}
This implies that
\begin{align}
I(U_j; Y_j|X_{2,j},Q_j)-I(U_j; Y_{2,j}|X_{2,j},Q_j) &\le I(U_j;
Y_j|X_{2,j},Q_j,Y_{2,j})\notag\\
&\le I(Q_j,U_j; Y_j|X_{2,j},Y_{2,j})\notag\\
&=0 \quad \text{for}~j\in \bar{\Ac} \label{eq:ap-dp3}
\end{align}
where the last equality of \eqref{eq:ap-dp3} follows from the Markov chain
relationship
\begin{align}
(Q_j,U_j) \rightarrow (X_{1,j},X_{2,j}) \rightarrow (Y_{2,j},X_{2,j})
\rightarrow Y_j \quad \text{for}~j\in \bar{\Ac}.
\end{align}
On the other hand, for $j\in \Av$, the subchannel satisfies
\begin{align}
p(y_j,y_{2,j}|x_{1,j},x_{2,j})=p(y_j|x_{1,j},x_{2,j})p(y_{2,j}|y_{j},x_{2,j}),
\quad \text{for}~j\in \Ac. \label{eq:app-dga}
\end{align}
Hence, we obtain
\begin{align}
I(U_j; Y_j|X_{2,j},Q_j)-I(U_j; Y_{2,j}|X_{2,j},Q_j) &\le I(U_j;
Y_j|X_{2,j},Q_j,Y_{2,j})\notag\\
&\le I(U_j,X_{1,j}; Y_j|X_{2,j},Q_j,Y_{2,j})\notag\\
&= I(X_{1,j}; Y_j|X_{2,j},Q_j,Y_{2,j}) \label{eq:ap-dp4-1}\\
&= I(X_{1,j}; Y_j,Y_{2,j}|X_{2,j},Q_j)-I(X_{1,j}; Y_{2,j}|X_{2,j},Q_j) \label{eq:ap-dp4-2}\\
&= I(X_{1,j}; Y_j|X_{2,j},Q_j)-I(X_{1,j}; Y_{2,j}|X_{2,j},Q_j) \quad
\text{for}~j\in \Ac \label{eq:ap-dp4}
\end{align}
where (\ref{eq:ap-dp4-1}) follows from the Markov chain relationship
\begin{align}
(Q_j,U_j,Y_{2,j}) \rightarrow (X_{1,j},X_{2,j}) \rightarrow Y_{j},
\end{align}
(\ref{eq:ap-dp4-2}) follows from the chain rule of mutual information, and
(\ref{eq:ap-dp4}) follows from the conditional degradedness \eqref{eq:app-dga}.
Now, substituting \eqref{eq:ap-dp3} and \eqref{eq:ap-dp4} into
\eqref{eq:ap-dp2}, we obtain the bound on $R_1$ given in \eqref{eq:pd-mac}.
This concludes the proof of the converse.

\subsection{Proof of Theorem~\ref{thm:pG-MAC}} \label{app:thm3}

By the {\it degraded, same-marginals} argument (see \cite{LP-IT08-MAC}), we
need to prove Theorem~\ref{thm:pG-MAC} only for the channel defined by
(\ref{eq:spGMAC-1})-(\ref{eq:spGMAC-2}).

\subsubsection*{Achievability} The achievability follows by applying Theorem~\ref{thm:pd}
and choosing the input distribution as follows
\begin{align}
Q_j&=\text{constant},~X_{2,j}\sim \Nc(0, p_{2,j}),\notag\\
X'_{1,j}&\sim \Nc(0, (1-\alpha_j) p_{1,j}),~X'_{1,j}~\text{is independent of}~X_{2,j}\notag\\
\text{and} \quad
X_{1,j}&=\sqrt{\frac{\alpha_jp_{1,j}}{p_{2,j}}}X_{2,j}+X'_{1,j}.
\end{align}
Moreover, by the fact $\alpha_j=1$ for $j\in \bar{\Ac}$, we obtain the secrecy
rate region $\Cc_s^{\rm [G]}$ is achievable.

\subsubsection*{Converse}

Here, we derive a tight upper bound on the achievable weighted sum rate
\begin{align*}
R_0+\gamma_1 R_1
\end{align*}
using Theorem~\ref{thm:pd} as the starting point. Since a capacity region is
always convex (via a time-sharing argument), an exact characterization of all the
achievable weighted sum rates for all nonnegative $\gamma_1$ provides an exact
characterization of the entire secrecy capacity region. By
Theorem~\ref{thm:pd}, any achievable rate pair $(R_0,R_1)$ must satisfy:
\begin{align}
R_0+\gamma_1 R_1 &\le \sum_{j\in \Ac} \left[ I(Q_j,X_{2,j}; Y_j)+\gamma_1
I(X_{1,j}; Y_j|X_{2,j},Q_j)-\gamma_1 I(X_{1,j};Y_{2,j}|X_{2,j},Q_j)\right]
\notag
\\&\quad + \sum_{j\in \bar{\Ac}}I(X_{1,j},X_{2,j}; Y_j).
\end{align}

For the subchannel $j\in \bar{\Ac}$, we are concerned only with the term
\begin{align}
I(X_{1,j},X_{2,j}; Y_j). \label{eq:ap3-T1}
\end{align}
The maximum-entropy theorem \cite{Cover} implies that (\ref{eq:ap3-T1}) is
maximized when $X_{1,j}$ and $X_{2,j}$ are jointly Gaussian with variance
$p_{1,j}$ and $p_{2,j}$ repetitively, and are aligned, i.e.,
$X_{1,j}=\sqrt{p_{1,j}/p_{2,j}}X_{2,j}$. Hence, we have
\begin{align}
I(X_{1,j},X_{2,j}; Y_j)\le \frac{1}{2}\log\left(1+\frac{p_{1,j}+p_{2,j}+2\sqrt{
p_{1,j}p_{2,j}}}{\nu_j}\right) \quad \text{for}~j\in \bar{\Ac}.
\label{eq:ap3-ro1}
\end{align}

For the subchannel $j\in \Ac$, we focus on the term
\begin{align*}
 I(Q_j,X_{2,j}; Y_j)+\gamma_1  I(X_{1,j}; Y_j|X_{2,j},Q_j)-\gamma_1
I(X_{1,j};Y_{2,j}|X_{2,j},Q_j).
\end{align*}
Based on the channel model defined in (\ref{eq:spGMAC-1})-(\ref{eq:spGMAC-2}),
we have
\begin{align}
 &I(Q_j,X_{2,j}; Y_j)+\gamma_1  I(X_{1,j}; Y_j|X_{2,j},Q_j)-\gamma_1
I(X_{1,j};Y_{2,j}|X_{2,j},Q_j) \notag\\
&= h(Y_j)+(\gamma_1-1)h(Y_j|X_{2,j},Q_j)-\gamma_1
h(Y_{2,j}|X_{2,j},Q_j)+\frac{\gamma_1}{2}\log\frac{\mu_j}{\nu_j}.
\label{eq:ap3-T2}
\end{align}
Now, we consider the following two cases.

\noindent{\it Case 1:} $\gamma_1\le 1$. In this case, note that
\begin{align}
h(Y_j|X_{2,j},Q_j) &\ge h(Y_j|X_{1,j},X_{2,j},Q_j)=\frac{1}{2}\log 2\pi e
\nu_j \notag\\
h(Y_j|X_{2,j},Q_j) &\ge h(Y_{2,j}|X_{1,j},X_{2,j},Q_j)=\frac{1}{2}\log 2\pi e
\mu_j \notag\\
\text{and}\qquad\qquad\qquad h(Y_j) &\le
\frac{1}{2}\log\left(p_{1,j}+p_{2,j}+2\sqrt{ p_{1,j}p_{2,j}}+\nu_j\right).
\end{align}
Hence, we have
\begin{align}
 &I(Q_j,X_{2,j}; Y_j)+\gamma_1  I(X_{1,j}; Y_j|X_{2,j},Q_j)-\gamma_1
I(X_{1,j};Y_{2,j}|X_{2,j},Q_j) \notag\\
&\le \frac{1}{2}\log\left(1+\frac{p_{1,j}+p_{2,j}+2\sqrt{
p_{1,j}p_{2,j}}}{\nu_j}\right) \quad \text{for}~j\in \Ac~\text{and}~\gamma_1\le
1. \label{eq:ap3-ro2}
\end{align}
This result implies that when the weight of the confidential-message rate is
less than the weight of the common-message rate, the optimum solution is to
allocate all possible power to transmit the common message.

\noindent{\it Case 2:} $\gamma_1> 1$. Without loss of generality, we assume
that the conditional covariance of $X_{1,j}$ given $(X_{2,j},Q_j)$ is given by
\begin{align}
\Cov(X_{1,j}|X_{2,j},Q_j) = \rho_j p_{1,j}
\end{align}
where $0\le \rho_j \le 1$. By applying the extremal inequality
\cite[Theorem~8]{LiuTie:it:07}, we have
\begin{align}
(\gamma_1-1)h(Y_j|X_{2,j},Q_j)-\gamma_1 h(Y_{2,j}|X_{2,j},Q_j) \le
\frac{\gamma_1-1}{2}\log 2 \pi e \left(\rho_j
p_{1,j}+\nu_j\right)-\frac{\gamma_1}{2}\log 2 \pi e \left(\rho_j
p_{1,j}+\mu_j\right). \label{eq:ap3-ca1}
\end{align}
Moreover, for a given $\rho_j$,
\begin{align}
h(Y_j) &\le \frac{1}{2}\log\left(p_{1,j}+p_{2,j}+2\sqrt{
(1-\rho_j)p_{1,j}p_{2,j}}+\nu_j\right). \label{eq:ap3-ca2}
\end{align}
Substituting (\ref{eq:ap3-ca1}) and (\ref{eq:ap3-ca2}) into (\ref{eq:ap3-T2}),
we obtain
\begin{align}
 &I(Q_j,X_{2,j}; Y_j)+\gamma_1  I(X_{1,j}; Y_j|X_{2,j},Q_j)-\gamma_1
I(X_{1,j};Y_{2,j}|X_{2,j},Q_j) \notag\\
&\le \max_{0\le \rho_j \le 1} \left[
\frac{1}{2}\log\left(1+\frac{p_{1,j}+p_{2,j}+2\sqrt{
(1-\rho_j)p_{1,j}p_{2,j}}}{\nu_j}\right) \right.\notag\\
&\qquad~~ \left. + \frac{\gamma_1-1}{2}\log 2 \pi e \left(1+\frac{\rho_j
p_{1,j}}{\nu_j}\right) -\frac{\gamma_1}{2}\log 2 \pi e \left(1+\frac{\rho_j
p_{1,j}}{\mu_j}\right)\right] \notag\\
&=\max_{0\le \alpha_j \le 1}
\left[\frac{\gamma_1}{2}\log\left(1+\frac{(1-\alpha_j)p_{1,j}}{\nu_j}\right)-
\frac{\gamma_1}{2}\log\left(1+\frac{(1-\alpha_j)p_{1,j}}{\mu_j}\right)\right.
\notag\\
&\qquad~~ \left.+
\frac{1}{2}\log\left(1+\frac{\alpha_jp_{1,j}+p_{2,j}+2\sqrt{\alpha_j
p_{1,j}p_{2,j}} }{(1-\alpha_j)p_{1,j}+\nu_j}\right)\right] \quad
\text{for}~j\in \Ac~\text{and}~\gamma_1> 1. \label{eq:ap3-ro3}
\end{align}
Finally, combining (\ref{eq:ap3-ro1}), (\ref{eq:ap3-ro2}) and
(\ref{eq:ap3-ro3}), we complete the converse proof.

\subsection{Proof of Theorem~\ref{thm:op1}} \label{app:op}

We need fine the optimal $\underline{p}^{\star}\in \Pc$ that maximizes
\begin{align}
 R_0+\gamma_1 R_1
\end{align}
where $\gamma_1\ge 0$. The Lagrangian is given by
\begin{align}
\Lc &= \sum_{j\in \Ac}
\left[\frac{1}{2}\log\left(1+\frac{a_j+p_{2,j}+2\sqrt{a_jp_{2,j}}
}{b_j+\nu_j}\right) + \frac{\gamma_1 }{2}\log\left(1+\frac{b_j}{\nu_j}\right)-
\frac{\gamma_1 }{2}\log\left(1+\frac{b_j}{\mu_j}\right)\right]
\notag\\
&\qquad + \sum_{j\in \bar{\Ac}}
\frac{1}{2}\log\left(1+\frac{a_j+p_{2,j}+2\sqrt{
a_jp_{2,j}}}{\nu_j}\right)-\lambda_1\left[\sum_{j\in\Ac}
(a_j+b_j)+\sum_{j\in\bar{\Ac}} a_j\right]-\lambda_2 \sum_{j=1}^{L} p_{2,j}
\label{eq:Lag-1}
\end{align}
where $\lambda_1$ and $\lambda_2$ are Largrange multiplier.


For $j\in\bar{\Ac}$, $(a_j^{\star}, p_{2,j}^{\star})$ needs to maximize the
following $\Lc_j$,
\begin{align}
\Lc_j=\frac{1}{2}\log\left(1+\frac{a_j+p_{2,j}+2\sqrt{
a_jp_{2,j}}}{\nu_j}\right)-\lambda_1 a_j-\lambda_2 p_{2,j}. \label{eq:Lag-B1}
\end{align}
Taking derivative of the Lagrangian in (\ref{eq:Lag-B1}) over $a_j$ and
$p_{2,j}$, the KKT conditions can be written as follows:
\begin{align}
\frac{1}{2\ln2}\frac{\theta_{1,j}(a_j,p_{2,j})}{\sqrt{a_j}} &=\lambda_1\notag\\
\text{and} \qquad
\frac{1}{2\ln2}\frac{\theta_{1,j}(a_j,p_{2,j})}{\sqrt{p_{2,j}}} &=\lambda_2
\label{eq:KKT-B1}
\end{align}
where
\begin{align}
\theta_{1,j}(a_j,p_{2,j})&=\frac{\sqrt{a_j}+\sqrt{p_{2,j}}}{\nu_j+
a_j+p_{2,j}+2\sqrt{a_j p_{2,j}}}.
\end{align}
This implies that the pair $(a_j^{\star}, p_{2,j}^{\star})$ that optimizes
$\Lc_j$ must satisfy
\begin{align}
p_{2,j}^{\star}=\left(\frac{\lambda_1}{\lambda_2}\right)^2 a_j^{\star}.
\label{eq:KKT-B2}
\end{align}
Let us define
\begin{align}
\beta=\lambda_1/\lambda_2.
\end{align}
On substituting (\ref{eq:KKT-B2}) into (\ref{eq:Lag-B1}), we obtain that
\begin{align}
\Lc_j &=\frac{1}{2}\log
\left[1+\frac{a_j(1+\beta)^2}{\nu_j}\right]-\lambda_1
a_j(1+\beta)\notag\\
&=\int_0^{a_j(1+\beta)^2} t_{1,j}(s) \,ds \label{eq:Lag-B2}
\end{align}
where
\begin{align}
t_{1,j}(s)=\frac{1}{(2\ln 2)}\frac{1}{(\nu_j+s)}-\frac{\lambda_1}{1+\beta}.
\label{eq:def-t1}
\end{align}
We define $s_{1,j}$ to be the root of the equation $t_{1,j}(s)=0$, i.e.,
\begin{align}
s_{1,j}&=\frac{1+\beta}{2\lambda_1\ln 2}-\nu_j\notag\\
&=\frac{\lambda_1+\lambda_2}{2\lambda_1\lambda_2\ln 2}-\nu_j. \label{eq:def-s1}
\end{align}
Hence, we obtain, for $j\in\bar{\Ac}$,
\begin{align}
a_j^{\star}&=\frac{1}{(1+\beta)^2}(s_{1,j})^{+}\notag\\
&=\frac{\lambda_2^2}{(\lambda_1+\lambda_2)^2}\left(\frac{\lambda_1+\lambda_2}{2\lambda_1\lambda_2\ln
2}-\nu_j\right)^{+}
\end{align}
and
\begin{align}
p_{2,j}^{\star}&=\beta^2 a_j^{\star}\notag\\
&=\frac{\lambda_1^2}{(\lambda_1+\lambda_2)^2}\left(\frac{\lambda_1+\lambda_2}{2\lambda_1\lambda_2\ln
2}-\nu_j\right)^{+}.
\end{align}


For $j\in\Ac$, $(a_j^{\star}, b_j^{\star}, p_{2,j}^{\star})$ needs to maximize
the following $\Lc_j$:
\begin{align}
\Lc_j&=\frac{1}{2}\log\left(1+\frac{a_j+p_{2,j}+2\sqrt{a_jp_{2,j}}
}{b_j+\nu_j}\right)+\frac{\gamma_1 }{2}\log\left(1+\frac{b_j}{\nu_j}\right)-
\frac{\gamma_1 }{2}\log\left(1+\frac{b_j}{\mu_j}\right) -\lambda_1
(a_j+b_j)-\lambda_2 p_{2,j}. \label{eq:Lag-A1}
\end{align}
Taking derivative of the Lagrangian in (\ref{eq:Lag-A1}) over $a_j$ and
$p_{2,j}$, the KKT conditions can be written as follows:
\begin{align}
\frac{1}{2\ln2}\frac{\theta_{2,j}(a_j,b_j,p_{2,j})}{\sqrt{a_j}} &=\lambda_1\notag\\
\text{and} \qquad
\frac{1}{2\ln2}\frac{\theta_{2,j}(a_j,b_j,p_{2,j})}{\sqrt{p_{2,j}}} &=\lambda_2
\label{eq:KKT-A1}
\end{align}
where
\begin{align}
\theta_{2,j}(a_j,b_j,p_{2,j})&=\frac{\sqrt{a_j}+\sqrt{p_{2,j}}}{\nu_j+
a_j+b_j+p_{2,j}+2\sqrt{a_j p_{2,j}}}.
\end{align}
This implies that the pair $(a_j^{\star}, p_{2,j}^{\star})$ that optimizes
$\Lc_j$ must satisfy
\begin{align}
p_{2,j}^{\star}=\left(\frac{\lambda_1}{\lambda_2}\right)^2a_j^{\star}=\beta^2 a_j^{\star}.
\label{eq:KKT-A2}
\end{align}
On substituting (\ref{eq:KKT-A2}) into (\ref{eq:Lag-A1}), we obtain that
\begin{align}
\Lc_j &=\frac{1}{2}\log
\left[1+\frac{a_j(1+\beta)^2}{b_j+\nu_j}\right]+\frac{\gamma_1
}{2}\log\left(1+\frac{b_j}{\nu_j}\right)- \frac{\gamma_1
}{2}\log\left(1+\frac{b_j}{\mu_j}\right)-\lambda_1 [a_j(1+\beta)+b_j]
\notag\\
&=\int_{b_j}^{b_j+a_j(1+\beta)^2} t_{1,j}(s) \,ds+\int_0^{b_j}
t_{2,j}(s)\,ds
\notag\\
&\le \int_0^{\infty}\left(\max\{t_{1,j}(s), t_{2,j}(s)\}
\right)^{+}\,ds\label{eq:Lag-A2}
\end{align}
where $t_{1,j}(s)$ is defined in (\ref{eq:def-t1}) and
\begin{align}
t_{2,j}(s)&=\frac{\gamma_1}{2\ln
2}\left(\frac{1}{\nu_j+s}-\frac{1}{\mu_j+s}\right)-\lambda_1. \label{eq:def-t2}
\end{align}

Next, we will derive $(a_j^{\star},b_j^{\star},p_{2,j}^{\star})$ that achieves
the upper bound on $\Lc_j$ in (\ref{eq:Lag-A2}). We consider the point of
intersection between $t_{1,j}(s)$ and $t_{2,j}(s)$. By using the definitions of
$t_{1,j}(s)$ in (\ref{eq:def-t1}) and $t_{2,j}(s)$ in (\ref{eq:def-t2}), the
point of intersection must satisfy
\begin{align}
\frac{1}{2\ln
2}\frac{1}{\nu_j+s}-\frac{\lambda_1}{1+\beta}=\frac{\gamma_1}{2\ln
2}\left(\frac{1}{\nu_j+s}-\frac{1}{\mu_j+s}\right)-\lambda_1,
\end{align}
i.e.,
\begin{align}
s^2+\left(\mu_j+\nu_j+\frac{1}{\omega}\right)s
+\left[\mu_j\nu_j-\frac{\gamma_1(\mu_j-\nu_j)-\mu_j}{\omega}\right]=0
\label{eq:Int-1}
\end{align}
where
\begin{align}
\omega&=(2\lambda_1\ln2)\frac{\beta}{1+\beta}\notag\\
&=(2\ln2)\frac{\lambda_1^2}{\lambda_1+\lambda_2}.
\end{align}
In the following, we consider two cases based on the relationship between
$\omega$ and $(\gamma_1(\mu_j-\nu_j)-\mu_j)/(\mu_j\nu_j)$.

\subsubsection{$\omega \ge \frac{\gamma_1(\mu_j-\nu_j)-\mu_j}{\mu_j\nu_j}$}
In this case, (\ref{eq:Int-1}) implies that the point of intersection between
$t_{1,j}(s)$ and $t_{2,j}(s)$ is either zero or negative. Moreover, it is easy
to see, for $s\ge 0$,
\begin{align}
t_{1,j}(s)-t_{2,j}(s)=\frac{(\nu_j+s)(\mu_j+s)\omega-[\gamma_1(\mu_j-\nu_j)-(\mu_j+s)]}
{(2\ln2)(\nu_j+s)(\mu_j+s)} \ge 0.
\end{align}
Hence, the upper bound on $\Lc_j$ in (\ref{eq:Lag-A2}) is achieved by
$b_j^{\star}=0$,
\begin{align}
a_j^{\star}
&=\frac{1}{(1+\beta)^2}(s_{1,j})^{+}\notag\\
&=\frac{\lambda_2^2}{(\lambda_1+\lambda_2)^2}\left(\frac{\lambda_1+\lambda_2}{2\lambda_1\lambda_2\ln
2}-\nu_j\right)^{+}
\end{align}
and
\begin{align}
p_{2,j}^{\star}
&=\beta^2 a_j^{\star}\notag\\
&=\frac{\lambda_1^2}{(\lambda_1+\lambda_2)^2}\left(\frac{\lambda_1+\lambda_2}{2\lambda_1\lambda_2\ln
2}-\nu_j\right)^{+}
\end{align}
where $s_{1,j}$ is defined in (\ref{eq:def-s1}).

\subsubsection{$\omega < \frac{\gamma_1(\mu_j-\nu_j)-\mu_j}{\mu_j\nu_j}$}

In this case, (\ref{eq:Int-1}) implies that, for $s>0$, $t_{1,j}(s)$ and
$t_{2,j}(s)$ intersect only once at
\begin{align}
\phi_j
=-\frac{1}{2}\left(\mu_j+\nu_j+\frac{1}{\omega}\right)+\frac{1}{2}\sqrt{\left(\mu_j+\nu_j+
\frac{1}{\omega}\right)^2-4\left[\mu_j\nu_j-\frac{\gamma_1(\mu_j-\nu_j)-\mu_j}{\omega}\right]}.
\end{align}
Moreover, it is easy to see that $t_{1,j}(0)<t_{2,j}(0)$. Hence, we have
\begin{align}
t_{1,j}(s)&<t_{2,j}(s)\qquad \text{for}~~ 0\le s <\phi_j\notag\\
\text{and}\qquad t_{1,j}(s)&\ge t_{2,j}(s)\qquad \text{for}~~ s \ge \phi_j.
\end{align}
Let $s_{2,j}$ denote the largest root of $t_{2,j}(s)=0$, i.e.,
\begin{align}
s_{2,j}&=\frac{1}{2}\left[\sqrt{(\mu_j-\nu_j)\left(\mu_j-\nu_j+\frac{2\gamma_1}{\lambda_1\ln
2}\right)}-(\mu_j+\nu_j)\right].
\end{align}
The optimal $(a_j^{\star},b_j^{\star},p_{2,j}^{\star})$ depends on the values
$t_{2,j}(0)$, $s_{1,j}$ and $\phi_j$, and falls into the following three
possibilities.

\noindent {\bf(2.a)} If $t_{2,j}(0)<0$, then both $t_{1,j}(s)$ and $t_{2,j}(s)$ are negative
for $s\ge 0$ (since both $t_{1,j}(s)$ and $t_{2,j}(s)$ are decreasing functions
for $s\ge0$). Then, the upper bound on $\Lc_j$ in (\ref{eq:Lag-A2}) is achieved
by $b_j^{\star}=0$, $a_j^{\star}=0$ and $p_{2,j}^{\star}=0$.

\noindent {\bf(2.b)} If $t_{2,j}(0) \ge 0$ and  $s_{1,j}<\phi_j$, then the upper bound
on $\Lc_j$ in (\ref{eq:Lag-A2}) is achieved by $b_j^{\star}=s_{2,j}$,
$a_j^{\star}=0$ and $p_{2,j}^{\star}=0$.

\noindent {\bf(2.c)} If $t_{2,j}(0) \ge 0$ and  $s_{1,j}\ge \phi_j$, then the upper
bound on $\Lc_j$ in (\ref{eq:Lag-A2}) is achieved by $b_j^{\star}=\phi_j$,
\begin{align}
a_j^{\star}&=\frac{1}{(1+\beta)^2}\left(s_{1,j}-\phi_j\right)
\quad \text{and} \quad
p_{2,j}^{\star}=\frac{\beta^2}{(1+\beta)^2}\left(s_{1,j}-\phi_j\right).
\end{align}

Combing the cases (2.a), (2.b) and (2.c), we obtain
\begin{align}
a_j^{\star}&
=\frac{\lambda_2^2}{(\lambda_1+\lambda_2)^2}\left(s_{1,j}-\phi_j\right)^{+}\notag\\
b_j^{\star}&=\left( \min\left[\phi_j,\, s_{2,j} \right]\right)^{+} \notag\\
\text{and}\qquad
p_{2,j}^{\star}&=\frac{\lambda_1^2}{(\lambda_1+\lambda_2)^2}\left(s_{1,j}-\phi_j\right)^{+}.
\end{align}

Finally, the Lagrange parameters $\lambda_1\ge 0$ and $\lambda_2\ge 0$ are
chosen to satisfy the power constraint (\ref{eq:pow1}).

\bibliographystyle{IEEEtran}
\bibliography{secrecy}

\end{document}